\documentclass[prd,preprint,superscriptaddress,showpacs,byrevtex]{revtex4}
\usepackage{epsfig}
\newcommand{\be}{\begin{equation}}
\newcommand{\ee}{\end{equation}}
\newcommand{\bea}{\begin{eqnarray}}
\newcommand{\eea}{\end{eqnarray}}
\begin{document}
%
%\begin{flushright}
%YITP-SB-06-xx
%\end{flushright}
%
\title{\noindent Renormalization Group Equation and QCD Coupling Constant in the Presence of SU(3)
Chromo-Electric Field }
\author{Gouranga C. Nayak} \email{nayak@physics.arizona.edu}
\affiliation{ Department of Physics, University of Arizona, Tucson, AZ 85721, USA }
\date{\today}
\begin{abstract}

We solve renormalization group equation in QCD in the presence of SU(3) constant chromo-electric
field $E^a$ with arbitrary color index $a$=1,2,...8 and find that the QCD coupling constant $\alpha_s$
depends on two independent casimir/gauge invariants $C_1=[E^aE^a]$ and $C_2=[d_{abc}E^aE^bE^c]^2$
instead of one gauge invariant $C_1=[E^aE^a]$. The $\beta$ function is derived from the
one-loop effective action. This coupling constant may be useful to study hadron formation from color
flux tubes/strings at high energy colliders and to study quark-gluon plasma formation at RHIC and LHC.

\end{abstract}
\maketitle
%\narrowtext

\newpage

\section{Introduction}

Although quantum chromodynamics describes the interaction among quarks and gluons,
the classical color field is used in many experimental situations, especially
to model the non-perturbative physics. Quark and gluon production from strong
chromo-electric field via Schwinger mechanism \cite{peter,schw,he,cash} itself is
a non-perturbative effect. This mechanism is often used in PYTHIA generator
\cite{pythia} to study low $p_T$ hadron production at collider experiments.
At RHIC and LHC heavy-ion colliders the classical color field play an important
role to study production of quark-gluon plasma \cite{qgp1}. Color glass condensate
provide a natural framework to determine the initial condition on the classical
color field at RHIC and LHC \cite{larry}.

In these situations it is necessary to know how QCD
coupling constant depends on SU(3) color field. In this paper we
solve the renormalization group equation in QCD in the presence of
SU(3) constant chromo-electric field $E^a$ with arbitrary color index
$a$=1,2,...8. Using background field method in QCD we derive
$\beta$ function from the one loop effective action of quark and gluon
in the presence of constant chromo-electric field $E^a$.
Using these two facts we determine the exact dependence of the QCD coupling
constant $\alpha_s$ on chromo-electric field $E^a$ in SU(3).

The paper is organized as follows. In section II we solve renormalization
group equation in QCD in the presence of SU(3) chromo-electric field. In
section III we derive $\beta$ function from one-loop effective action. In
section IV we discuss the dependence of QCD coupling constant on the second
casimir invariant in SU(3). We present our conclusions in section V.

\section{ Renormalization Group Equation in QCD in SU(3) Chromo-Electric Field }

In the background field method of QCD \cite{thooft,abbott} the total gauge field is the sum of
classical background field $A_\mu^a$ and quantum gluon field $Q_\mu^a$. The gauge fixing
term depends on the background field $A_\mu^a$. As pointed out in \cite{abbott} it is
not necessary to renormalize the quantum gluon fields $Q_\mu^a$ and the ghost fields.
Gauge fixing parameter renormalization is also not necessary because the physical result is
gauge fixing parameter independent. Hence the background field $A_\mu^a$
and coupling constant $g$ need to be renormalized.
We define the bare quantities in terms of the renormalized one as follows \cite{abbott}
\bea
 A_{\mu }^a = Z_A^{-\frac{1}{2}}~A_{\mu r}^a,~~~~~~~~~~~~~ g = Z_g~g_r.
\label{rr}
\eea
This gives
\bea
F_{\mu \nu }^a[A] =
Z_A^{-\frac{1}{2}}[\partial_\mu A_{\nu r}^a
-\partial_\nu A_{\mu r}^a +Z_A^{-{\frac{1}{2}}}Z_g~gf^{abc} ~A_{\mu r}^b ~A_{\nu r}^c].
\eea
Since $F_{\mu \nu }^a[A]$ transforms covariantly with respect to gauge
transformation we find from the above equation
\bea
Z_g=Z_A^{\frac{1}{2}},~~~~~~~~~~~~~{\rm which~~ gives},~~~~~~~~~~~~~\beta =  -g\gamma,
\label{bg}
\eea
where we have used
\bea
\beta = -g \mu \frac{ \partial}{\partial \mu} {\rm Ln} Z_g,~~~~~~~~~~~~~~{\rm and}~~~~~~~~~~~~~ \gamma= \frac{1}{2} \mu \frac{ \partial}{\partial \mu} {\rm Ln} Z_A.
\eea

The renormalization group equation for the effective action $\Gamma $ can be written as \cite{weinberg}
\bea
[ \mu^2 \frac{\partial}{\partial \mu^2}
+ \beta \frac{\partial}{\partial g}
+ \gamma \int d^4x A^a_\mu \frac{\delta}{\delta A^a_\mu} ] \Gamma =0.
\label{wein}
\eea

The effective action may be written in terms of the 1PI Green's function via
\bea
\Gamma = \sum_n \frac{1}{n!} \int d^4x_1 .....d^4x_n
{\Gamma^{(n)}}^{a_1,....,a_n}_{\mu_1,....\mu_n}(x_1,...x_n)
A_{\mu_1}^{a_1}...... A_{\mu_n}^{a_n}.
\label{1pi}
\eea

Using eq. (\ref{1pi}) in (\ref{wein}) we find
\bea
[ \mu^2 \frac{\partial}{\partial \mu^2}
+ \beta \frac{\partial}{\partial g}
+ n\gamma ] \Gamma^{(n)} =0
\label{effn}
\eea
which is the familiar form of the renormalization group equation in QCD \cite{muta}.

There is another way to expand the effective action \cite{schw,weinberg}.
Instead of expanding in powers of $A_\mu^a$ one can expand in
powers of momentum. In coordinate space it has the form
\bea
\Gamma = \int d^4x {\cal L}=\int d^4x~[-V(A)+\frac{1}{2} (\partial_\mu A_\nu^a)^2 Z_A +....].
\label{el}
\eea
Hence we will use eq. (\ref{wein}) instead of (\ref{effn}) for the differential
equation of the renormalization group. In the presence of constant chromo-electric
field the one loop effective action for gluon
\bea
{\cal L}^{(1)} = \sum_{j=1}^3 {\cal L}^{(1)}_j
\label{sum}
\eea
depends on three independent gauge and Lorentz invariant
eigenvalues $\lambda_j$ of $f^{abc}E^c$ in SU(3)
\bea
\lambda_1^2=\frac{C_1}{2}[1-{\rm cos}\theta],~~~~~~~~\lambda_{2,3}^2=\frac{C_1}{2}[1+{\rm cos}(\frac{\pi}{3}\pm \theta)],~~~~~~~~{\rm cos}3\theta = -1+\frac{6C_2}{C_1^3}.
\label{eigen}
\eea
Hence the renormalized effective lagrangian density ${\cal L}_j$ depends on $\lambda_j$.
In terms of gauge and Lorentz invariant variables, the renormalization group equation
becomes
\bea
[ \mu^2 \frac{\partial}{\partial \mu^2}
+ \beta \frac{\partial}{\partial g}
+ \gamma \lambda_j \frac{\partial}{\partial \lambda_j} ] {\cal L}_j =0.
\label{renj}
\eea
In order to solve this differential equation we define dimensionless Lagrangian density
\bea
{\bar {\cal L}_j}= \frac{ { {\cal L}_j}}{\frac{1}{3} \lambda_j^2}
\label{ff}
\eea
which can only depend on the dimensionless quantity
\bea
t_j={\rm ln}~r_j={\rm ln}~\frac{g\lambda_j}{\mu^2}.
\label{tj}
\eea
We find from eq. (\ref{renj})
\bea
\frac{1}{{\bar {\cal L}_j}(t_j) }
\frac{\partial {\bar {\cal L}_j}(t_j) }{\partial t_j} +
\frac{{\bar \beta}}{{\bar {\cal L}_j}(t_j) } \frac{\partial
{\bar {\cal L}_j}(t_j,g) }{\partial g}=-2{\bar \gamma}
\label{rent}
\eea
where
\bea
{\bar \beta} = \frac{\beta}{(\gamma-1)},~~~~~~~~~~{\rm and}~~~~~~~~~~~~~ {\bar \gamma} = \frac{\gamma}{(\gamma-1)}.
\label{by}
\eea
Solving the differential eq. (\ref{rent}) we find:
\bea
{-\bar {\cal L}_j}(t_j)  ~ dt_j =
-\frac{{\bar {\cal L}_j}(t_j) }{\bar \beta}~dg =
\frac{d{{\bar {\cal L}_j}(t_j) }}{2{\bar \gamma}}
\label{dif}
\eea
which gives
\bea
\frac{dg}{dt_j}={\bar \beta }.
\label{bb}
\eea
In this paper we consider $\alpha_s$ at the one loop level and take the $\beta$ function of the form
\bea
{\bar \beta}(g)=-{\bar \beta_0} g^3.
\label{bb0}
\eea
We find from eq. (\ref{bb}) the QCD coupling constant
\bea
\alpha_s(\lambda_j)=\frac{g^2(t_j)}{4\pi} = \frac{g^2}{4\pi [1+ 2 {\bar \beta_0} g^2 t_j]}= \frac{\alpha_s}{[1+  4\pi {\bar \beta_0} \alpha_s {\rm Log}(\frac{g^2\lambda_j^2}{\mu^4})]}=\frac{1}{4 \pi {\bar \beta_0} {\rm Log}(\frac{g^2\lambda_j^2}{\Lambda^4})}
\label{llg2}
\eea
where
\bea
\Lambda = \mu ~e^{(-1/(4{\bar \beta_0} g^2))},~~~~~~~~~~~~~~~~\alpha_s=\frac{g^2}{4\pi}.
\label{scale}
\eea
For the quark case the $\beta$ functions are different and the eigenvalues are in fundamental representation
of SU(3). Three independent gauge and Lorentz invariant eigenvalues of $T^a_{ij}E^a$ for the quark case are
given by
\bea
\lambda_1=\sqrt{\frac{C_1}{3}}{\rm cos}\theta,~~~~~~~~\lambda_{2,3}=\sqrt{\frac{C_1}{3}}{\rm cos}(\frac{2\pi}{3}\pm \theta),~~~~~~~~{\rm cos}^23\theta = \frac{3C_2}{C_1^3}.
\label{eigeng}
\eea

All now remains is to find the $\beta$ functions from one-loop effective action which we will derive
in the next section.

\section{$\beta$ function in QCD from one-loop effective action}

The one-loop effective lagrangian density for gluon in the presence of SU(3) constant chromo-electric
field $E^a$ with $a$=1,2,...8 is given by \cite{peter}
\bea
{\cal L}^{(1)}_j = \frac{1}{8 \pi^2} \int_0^\infty \frac{ds}{s^2}
[g\lambda_j \frac{{\rm cos}2g\lambda_j s}{{\rm sin}g\lambda_js} -\frac{1}{s}].
\label{lj2}
\eea
Expanding ${\rm sin}$ and ${\rm cos}$ functions we get
\bea
{\cal L}^{(1)}_j = \frac{1}{8 \pi^2} \int_0^\infty \frac{ds}{s^2}
[-\frac{11 g^2\lambda_j^2s}{6} +\frac{39 g^4\lambda_j^4 }{240}s^3+...].
\label{lj3}
\eea
Since $s$ has dimension of length, the ultra violate divergence occurs at $s \rightarrow 0$ which
leads to charge renormalization \cite{schw}. The ultra violate divergent term in eq. (\ref{lj3}) is $\frac{1}{8 \pi^2} \int_0^\infty \frac{ds}{s} \frac{11 g^2\lambda_j^2}{6}$. We write eq. (\ref{lj2}) as follows
\bea
{\cal L}^{(1)}_j = \frac{1}{8 \pi^2} \int_0^\infty \frac{ds}{s^2}
[g\lambda_j \frac{{\rm cos}2g\lambda_j s}{{\rm sin}g\lambda_js}+\frac{11 g^2\lambda_j^2s}{6} -\frac{1}{s}]
-\frac{1}{8 \pi^2} \int_0^\infty \frac{ds}{s} \frac{11 g^2\lambda_j^2}{6}.
\label{lj4}
\eea
The $s$ integration involving the square bracket term is now finite. We can obtain the
$\beta$ function from the coefficient of the divergent term by renormalization procedure by adding the
counter term $Z{\cal L}_0$. We put a cut-off for the infrared limit at
$s=s_0$ and by change to the dimensionless variable $s_j \rightarrow sg\lambda_j$. We find
\bea
{\cal L}^{(1)}_j = \frac{g^2\lambda_j^2}{8 \pi^2} \int_0^\infty \frac{ds_j}{s_j^2}
[\frac{{\rm cos}2s_j}{{\rm sin}s_j}+\frac{11}{6} -\frac{1}{s_j}]
-\frac{1}{8 \pi^2} \int_0^{g\lambda_j s_0} \frac{ds_j}{s_j} \frac{11 g^2\lambda_j^2}{6}.
\label{lj6}
\eea
The free lagrangian density is given by
\bea
{\cal L}_0 = \frac{E^aE^a}{2} = \frac{1}{3}\sum_{j=1}^3 \lambda_j^2 ~=~\sum_{j=1}^3 {\cal L}_0^j.
\label{l0g}
\eea
Adding the counter term ($Z{\cal L}_0^j$) to eq. (\ref{lj6}) the renormalized lagrangian density becomes
\bea
{\cal L}_j(t_j) =Z{\cal L}_0^j+{\cal L}^{(1)}_j.
\label{tl}
\eea
The renormalization condition is then given by
\bea
\frac{{\cal L}_j(t_j)}{{\cal L}_0^j}|_{t_j=0} =1.
\label{rng}
\eea
We find from the above equation
\bea
Z=1- 3 \frac{g^2}{8 \pi^2} \int_0^\infty \frac{ds_j}{s_j^2}
[\frac{{\rm cos}2s_j}{{\rm sin}s_j}+\frac{11}{6} -\frac{1}{s_j}]
+\frac{3}{8 \pi^2} \int_0^{g\mu^2 s_0} \frac{ds_j}{s_j} \frac{11 g^2}{6}
\label{zg}
\eea
which when used in eq. (\ref{tl}) gives the renormalized Lagrangian density
\bea
{\bar {\cal L}}_j= \frac{{\cal L}_j(t_j)}{{\cal L}_0^j}= 1 - \frac{11}{16 \pi^2} g^2 t_j.
\label{tl3g}
\eea

The $\beta$ function can be obtained from the renormalized Lagrangian density.
From eqs. (\ref{rent}), (\ref{ff}) and (\ref{rng}) we find
\bea
{\bar \gamma}=-\frac{1}{2}
\frac{\partial {{\bar {\cal L}}_j}(t_j) }{\partial t_j}|_{t_j=0}.
\label{rent1}
\eea
Using eqs. (\ref{by}) and (\ref{bg}) in the above equation we find
\bea
{\bar \beta}=\frac{g}{2}~\frac{\partial {{\bar {\cal L}}_j}(t_j) }{\partial t_j}|_{t_j=0}.
\label{betaf}
\eea

Using eq. (\ref{betaf}) we find from eq. (\ref{tl3g})
\bea
{\bar \beta} =- \frac{11}{32 \pi^2} g^3
\label{bbar}
\eea
which gives the $\beta$ function for a gluon loop
\bea
{\bar {\beta_0}}^g=\frac{11}{32 \pi^2}
\label{b0g}
\eea
where we have used eq. (\ref{bb0}).

The effective Lagrangian density for massless quark is given by \cite{peter}
\bea
{\cal L}^{(1)}_j = -\frac{1}{8 \pi^2} \int_0^\infty \frac{ds}{s^2}
[g\lambda_j \frac{{\rm cos}g\lambda_j s}{{\rm sin}g\lambda_js} -\frac{1}{s}].
\label{lj2q}
\eea
Expanding ${\rm sin}$ and ${\rm cos}$ functions we get
\bea
{\cal L}^{(1)}=- \frac{1}{8 \pi^2} \int_0^\infty \frac{ds}{s}
[\frac{ g^2\lambda_j^2}{3} +\frac{ g^4\lambda_j^4 }{20}s^2+...].
\label{qeff1}
\eea

The coefficient of the ultra violate divergent term (as $s \rightarrow 0$) is
$\frac{1}{3}$. The free Lagrangian density is given by
\bea
{\cal L}_0 = \frac{E^aE^a}{2} = \sum_{j=1}^3 \lambda_j^2 ~=~\sum_{j=1}^3 {\cal L}_0^j.
\label{l0q}
\eea
Carrying out similar renormalization analysis as for gluons we obtain the renormalized
Lagrangian density
\bea
{\bar {\cal L}}_j= 1 + \frac{1}{24 \pi^2} g^2 t_j
\label{tl3q}
\eea
which gives the $\beta$ function for a quark loop
\bea
{\bar {\beta_0}}^q=-\frac{1}{48 \pi^2}.
\label{b0q}
\eea

To summarize, eq. (\ref{llg2}) describes evolution of QCD
coupling constant in the presence of SU(3) constant chromo-electric field,
together with eqs. (\ref{b0g}) and (\ref{b0q}) as ${\bar \beta}_0$ functions
for a gluon and quark loop respectively.

\section{ QCD Coupling Constant and Second Casimir Invariant in SU(3) }

It can be seen that the two independent casimir invariants $C_1=[E^aE^a]$ and $C_2=[d_{abc}E^aE^bE^c]^2$
in SU(3) are gauge invariant with respect to the gauge transformation
\bea
(T^a E^a)' = U(T^a U^a) U^{-1},~~~~~~~~~~~~~~~~U=e^{iT^a \omega^a(x)}.
\label{gauge}
\eea
Let us denote the vector ${\vec E}$ in 8-dimensional color space in SU(3) with components $E_a=(E_1,~E_2,~.....,E_8)$.
It can be seen that while the first casimir invariant $C_1$ is independent of the direction of the vector ${\vec E}$,
the second casimir invariant $C_2$ may depend on the direction of ${\vec E}$ in 8-dimensional color space even if it is 
gauge invariant. This is because of presence of $d_{abc}$ whose components are not equal to each other \cite{pdg}. 
These two independent casimir invariants satisfy the limit
\bea
0 \le \frac{3C_2}{C_1^3} \le 1.
\label{c1c21}
\eea
Eq. (\ref{c1c21}) gives the limit
\bea
0 \le \theta \le \frac{2\pi}{3}
\label{gg}
\eea
when the $\lambda_j$'s are given by eq. (\ref{eigen}) in the adjoint representation of SU(3) and
\bea
\frac{\pi}{6} \le \theta \le \frac{\pi}{2}
\label{qq}
\eea
when the $\lambda_j$'s are given by eq. (\ref{eigeng}) in the fundamental representation of SU(3).

Since $\theta$ appears inside $log$ in $\alpha_s(\lambda_j)$ it is useful to check that this $\theta$ dependence is
under control for this one loop calculation. When the $\lambda_j$'s are given by eq. (\ref{eigen}),
it can be checked that the one-loop result of $\alpha_s(\lambda_j)$ in the maximum allowed range
$0 \le \theta \le \frac{2\pi}{3}$ (see eq. (\ref{gg})) is under control only for
asymptotically very large value of $C_1=[E^a E^a]$ (see below). Similarly, when the $\lambda_j$'s
are given by eq. (\ref{eigeng}), the one-loop result of $\alpha_s(\lambda_j)$ in the maximum allowed
range $\frac{\pi}{6} \le \theta \le \frac{\pi}{2}$ (see eq. (\ref{qq})) is under control only for
asymptotically very large value of $C_1=[E^a E^a]$.

For realistic values of $C_1=[E^aE^a]$, $g$ and $\Lambda$ at RHIC and LHC we can determine the range of $\theta$ for which
$0 < \alpha_s \le 1$ in this one-loop calculation of $\alpha_s$. For example, by using eq. (\ref{eigen}) we find from
$0 < \alpha_s \le 1$, the range
\bea
{\rm cos}^{-1} [ 1~-~\frac{2\Lambda^4}{g^2 C_1}e^{\frac{1}{4\pi{\bar \beta_0}}}]~\le~\theta ~\le ~-\pi/3 ~+~{\rm cos}^{-1}[\frac{2\Lambda^4}{g^2 C_1}e^{\frac{1}{4\pi{\bar \beta_0}}}~-~1].
\label{ntgl}
\eea
It can be seen that when $C_1=[E^aE^a]$ is asymptotically very large, say, $C_1 \rightarrow \infty$ we find
from eq. (\ref{ntgl}), $ 0 \le \theta \le \frac{2\pi}{3}$ which reproduces the maximum allowed range as given
by eq. (\ref{gg}). For realistic values of $C_1=[E^a E^a]$, $g$ and $\Lambda$ at RHIC and LHC the range in
eq. (\ref{ntgl}) may be be very close to the maximum allowed range in eq. (\ref{gg}).
For example, if we choose $\Lambda$=0.2 GeV, g=3 and $C_1=E^aE^a$ = 1000 GeV$^4$ (which may be a reasonable
value at LHC) we find from eq. (\ref{ntgl}), $0.0021 \le \theta \le 2.092$. This range of $\theta$ is very
close to the maximum allowed range $0 \le \theta \le \frac{2\pi}{3}$ or $0 \le \theta \le 2.094 $ as given by eq. (\ref{gg}).

Similarly, by using eq. (\ref{eigeng}) we find from $0 < \alpha_s \le 1$, the range
\bea
\frac{2\pi}{3} - {\rm cos}^{-1} [\sqrt{\frac{3\Lambda^4}{g^2 C_1}e^{\frac{1}{4\pi{\bar \beta_0}}}}]~\le~\theta ~\le ~
{\rm cos}^{-1} [\sqrt{\frac{3\Lambda^4}{g^2 C_1}e^{\frac{1}{4\pi{\bar \beta_0}}}}].
\label{ntql}
\eea
It can be seen that when $C_1=[E^aE^a]$ is asymptotically very large, say, $C_1 \rightarrow \infty$ we find
from eq. (\ref{ntql}), $ \frac{\pi}{6} \le \theta \le \frac{\pi}{2}$ which reproduces the maximum allowed range as given
by eq. (\ref{qq}). If we choose the previous values, $\Lambda$=0.2 GeV, g=3 and $C_1=E^aE^a$ = 1000 GeV$^4$ we find from
eq. (\ref{ntql}), $0.526 \le \theta \le 1.569$. This range of $\theta$ is very close to the maximum allowed range
$\frac{\pi}{6} \le \theta \le \frac{\pi}{2}$ or $0.5236 \le \theta \le 1.571 $ as given by eq. (\ref{qq}).

In Fig. 1 we present the result of $\alpha_s(\lambda_j)$ as function of $\theta$ for fixed values of
$C_1=[E_aE_a]$. We have used $g$=3 and $\Lambda$= 200 MeV. The scale $\lambda_j$'s are given by
eq. (\ref{eigen}). The range of $\theta$ in Fig. 1 is given by eq. (\ref{ntgl}). The upper, middle and
lower solid lines are the results of $\alpha_s(\lambda_1)$ for $C_1$ = 10, 1000 and 100000 GeV$^4$
respectively. The upper, middle and lower dotted lines are the results of $\alpha_s(\lambda_2)$ for
$C_1$ = 10, 1000 and 100000 GeV$^4$ respectively. The upper, middle and lower dashed lines are the
results of $\alpha_s(\lambda_3)$ for $C_1$ = 10, 1000 and 100000 GeV$^4$ respectively. It can be seen
from Fig. 1 that the $\theta$ dependence is under control for the entire range of $\theta$ as given
by eq. (\ref{ntgl}) which is very close to the actual maximum range $0 \le \theta \le \frac{2\pi }{3}$
in eq. (\ref{gg}).
\begin{figure}[htb]
\vspace{2pt}
%\centering{\rotatebox{270}{\epsfig{figure=sbcr.ps,height=7cm}}}
\centering{{\epsfig{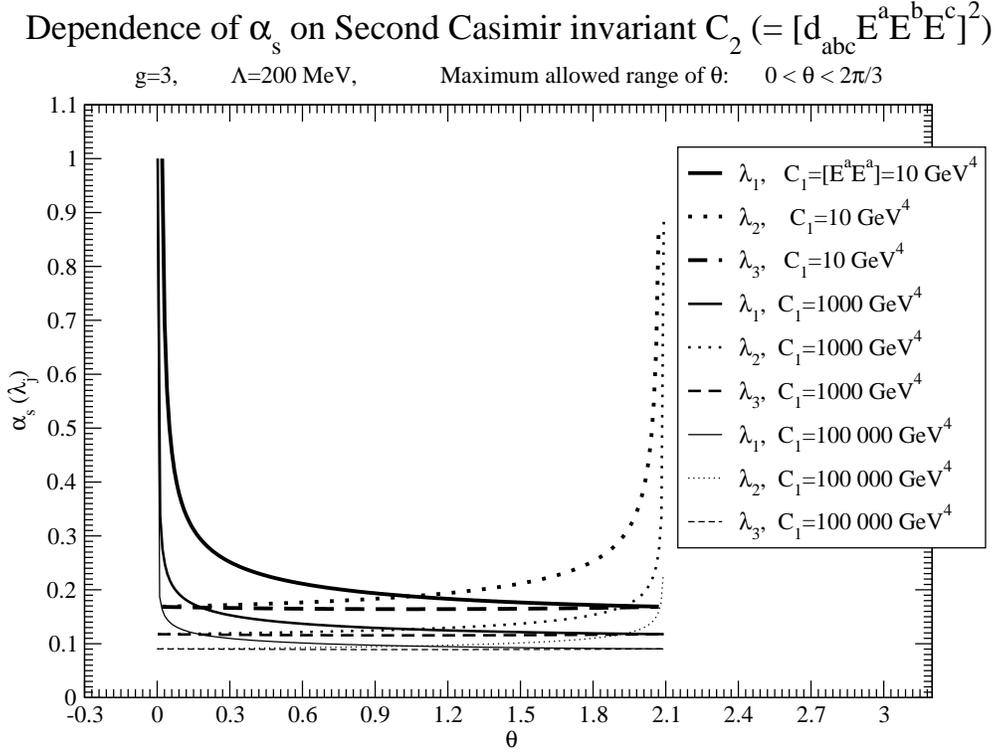}}}
\caption{ QCD coupling constant in the presence of SU(3) chromo-electric field as function $\theta$ for fixed values
of first casimir invariant $C_1=E^aE^a$. The $\lambda_j$'s used are from eq. (\ref{eigen}). Remember that
$0 \le \theta \le \frac{2\pi}{3}$ is the maximum range when the $\lambda_j$'s are given by eq. (\ref{eigen}).
}
\label{fig1}
\end{figure}

In Fig. 2 we present the result of $\alpha_s(\lambda_j)$ as function of $\theta$ for fixed values of
$C_1=[E_aE_a]$. We have used $g$=3 and $\Lambda$= 200 MeV. The scale $\lambda_j$'s are given by
eq. (\ref{eigeng}). The range of $\theta$ in Fig. 2 is given by eq. (\ref{ntql}). The upper, middle and
lower solid lines are the results of $\alpha_s(\lambda_1)$ for $C_1$ = 10, 1000 and 100000 GeV$^4$
respectively. The upper, middle and lower dotted lines are the results of $\alpha_s(\lambda_2)$ for
$C_1$ = 10, 1000 and 100000 GeV$^4$ respectively. The upper, middle and lower dashed lines are the
results of $\alpha_s(\lambda_3)$ for $C_1$ = 10, 1000 and 100000 GeV$^4$ respectively. It can be seen
from Fig. 2 that the $\theta$ dependence is under control for the entire range of $\theta$ as given
by eq. (\ref{ntql}) which is very close to the actual maximum range $\frac{\pi}{6} \le \theta \le \frac{\pi }{2}$
in eq. (\ref{qq}).

\begin{figure}[htb]
\vspace{2pt}
%\centering{\rotatebox{270}{\epsfig{figure=sbcr.ps,height=7cm}}}
\centering{{\epsfig{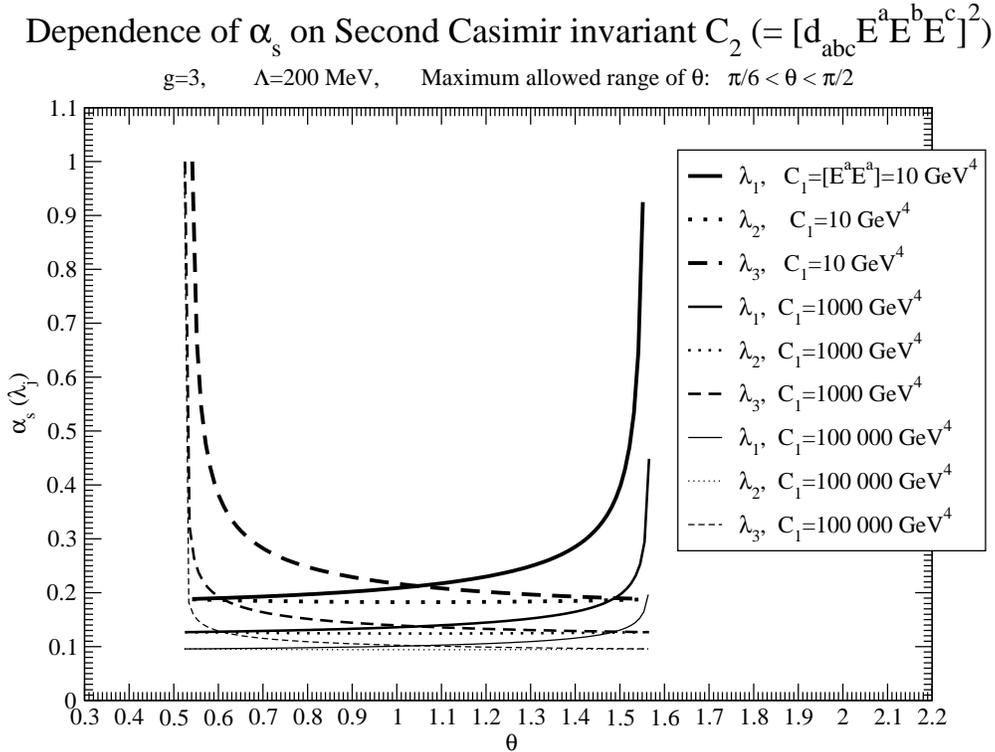}}}
\caption{QCD coupling constant in the presence of SU(3) chromo-electric field as function $\theta$ for fixed values
of first casimir invariant $C_1=E^aE^a$.The $\lambda_j$'s used are from eq. (\ref{eigeng}).
Remember that $\frac{\pi}{6} \le \theta \le \frac{\pi}{2}$ is the maximum
range when the $\lambda_j$'s are given by eq. (\ref{eigeng}).
}
\label{fig2}
\end{figure}

It can be seen from Fig. 1 and Fig. 2 that, for same values of $\theta$, the value of $\alpha_s$ decreases as
$C_1=[E^a E^a]$ increases which is consistent with asymptotic freedom.

In Fig. 3 we present the result of $\alpha_s(\lambda_j)$ as function of $C_1=[E_aE_a]$ for fixed values of
$\theta$. We have used $g$=3 and $\Lambda$= 200 MeV. The scale $\lambda_j$'s are given by
eq. (\ref{eigen}). The solid line is the result of $\alpha_s(\lambda_1)$ for $\theta ~=~\frac{\pi}{6}$.
The dotted line is the result of $\alpha_s(\lambda_2)$ for $\theta ~=~\frac{\pi}{3}$ and the dashed line
is the result of $\alpha_s(\lambda_3)$ for $\theta ~=~\frac{\pi}{2}$. In this figure we have chosen three different values
of $\theta$ which are within the maximum allowed range $0 \le \theta \le \frac{2\pi }{3}$ as given by eq. (\ref{gg}).
It can be seen from Fig. 3 that the $\theta$ dependence is under control in the entire range of $C_1=[E_aE_a]$.

\begin{figure}[htb]
\vspace{2pt}
%\centering{\rotatebox{270}{\epsfig{figure=sbcr.ps,height=7cm}}}
\centering{{\epsfig{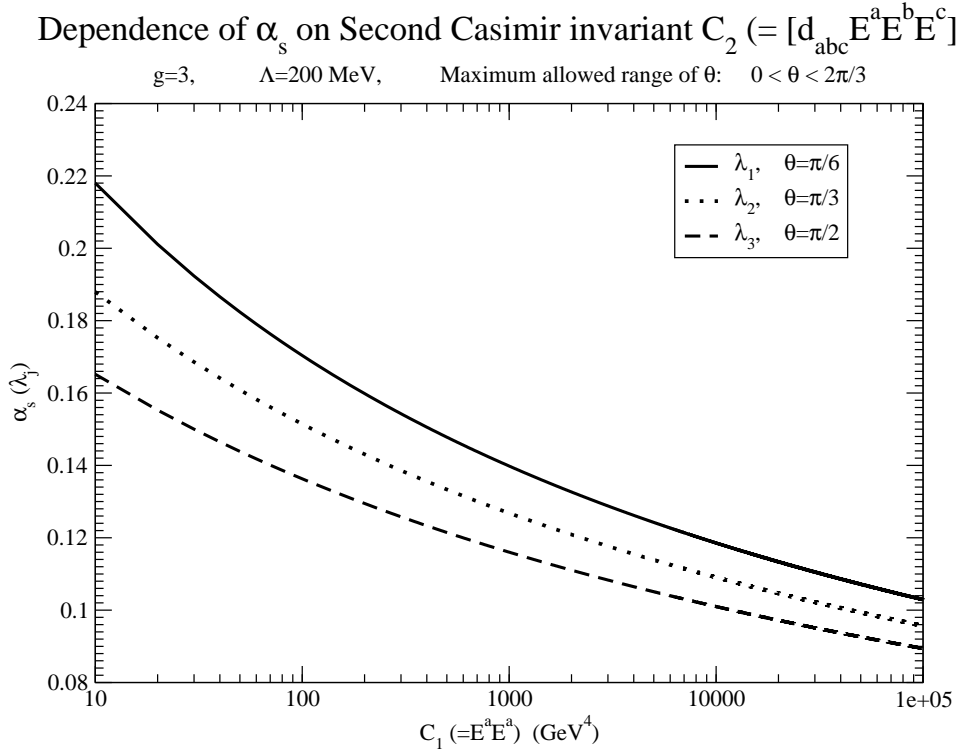}}}
\caption{ QCD coupling constant in the presence of SU(3) chromo-electric field as function of first casimir invariant
$C_1=[E^aE^a]$ for fixed values of $\theta$. The $\lambda_j$'s used are from eq. (\ref{eigen}).
Remember that $0 \le \theta \le \frac{2\pi}{3}$ is the maximum range when
the $\lambda_j$'s are given by eq. (\ref{eigen}).
}
\label{fig3}
\end{figure}

In Fig. 4 we present the result of $\alpha_s(\lambda_j)$ as function of $C_1=[E_aE_a]$ for fixed values of
$\theta$. We have used $g$=3 and $\Lambda$= 200 MeV. The scale $\lambda_j$'s are given by
eq. (\ref{eigeng}). The solid line is the result of $\alpha_s(\lambda_1)$ for $\theta ~=~\frac{5\pi}{12}$.
The dotted line is the result of $\alpha_s(\lambda_3)$ for $\theta ~=~\frac{\pi}{3}$ and the dashed line
is the result of $\alpha_s(\lambda_2)$ for $\theta ~=~\frac{\pi}{4}$. In this figure we have chosen three different values
of $\theta$ which are within the maximum allowed range $\frac{\pi}{6} \le \theta \le \frac{\pi }{2}$ as given by eq. (\ref{qq}).
It can be seen from Fig. 4 that the $\theta$ dependence is under control in the entire range of $C_1=[E_aE_a]$.

\begin{figure}[htb]
\vspace{2pt}
%\centering{\rotatebox{270}{\epsfig{figure=sbcr.ps,height=7cm}}}
\centering{{\epsfig{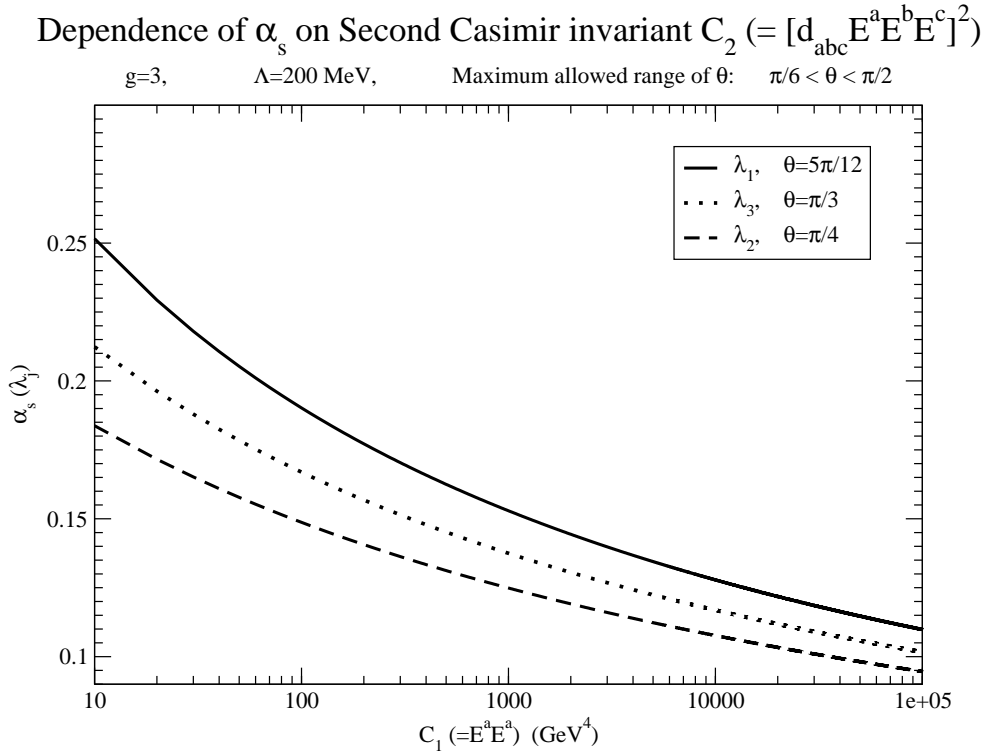}}}
\caption{ QCD coupling constant in the presence of SU(3) chromo-electric field as function of first casimir invariant
$C_1=[E^aE^a]$ for fixed values of $\theta$. The $\lambda_j$'s used are from eq. (\ref{eigen}).
Remember that $\frac{\pi}{6} \le \theta \le \frac{\pi}{2}$ is the maximum
range when the $\lambda_j$'s are given by eq. (\ref{eigeng}).
}
\label{fig4}
\end{figure}

It can be seen from Fig. 3 and Fig. 4 that the value of $\alpha_s$ decreases as $C_1=[E^a E^a]$ increases which is
consistent with asymptotic freedom.

\section{Conclusion}

We have solved renormalization group equation in QCD in the presence of SU(3)
constant chromo-electric field $E^a$ with arbitrary color index $a$=1,2,...8. Using background
field method in QCD we have obtained the $\beta$ function from one loop effective action of quark
and gluon in the presence of chromo-electric field $E^a$ in SU(3). Using these two facts we have
determined the exact dependence of the QCD coupling constant on $E^a$. We have found that the
renormalization scale of the QCD coupling constant $\alpha_s$ depends on two independent
casimir/gauge invariants $C_1=[E^aE^a]$ and $C_2=[d_{abc}E^aE^bE^c]^2$ instead of one gauge
invariant $C_1=[E^aE^a]$. These coupling constant may play an important role in the study of
production and equilibration of quark-gluon plasma from classical color field at RHIC and LHC.
This coupling constant may also play an important role to study low $p_T$ hadron production
at collider experiments using string breaking mechanism in the color flux-tube model.

\acknowledgments

This work was supported in part by Department of Energy under contracts
DE-FG02-91ER40664, DE-FG02-04ER41319 and DE-FG02-04ER41298.

\end{document}